\documentclass[12pt]{mypaper}
\pdfoutput=1
\usepackage{graphicx}
\usepackage{rotate}
\usepackage{rotating}
\usepackage{relsize}
\usepackage{lineno}
\usepackage{slashed}
\usepackage{url}
\usepackage{amsmath}
\usepackage{longtable}
\usepackage{setspace} 
\usepackage[font=small]{caption}
\usepackage[bottom]{footmisc}
    
\usepackage{latexsym}
\usepackage{amssymb}
\usepackage{epsfig,amsmath,graphics}
\usepackage{color}
\usepackage{dcolumn}
\usepackage{slashed}
\usepackage{comment,latexsym} 
\usepackage{bm}
\usepackage{verbatim}
\usepackage{tabularx}
\usepackage{dcolumn}

\usepackage{setspace}
\usepackage{xspace}
\usepackage{bigints}
\usepackage{multirow}
\usepackage{cite}
\usepackage{lineno}

\definecolor{colorLink}{rgb}{0.7,0,0}
\definecolor{colorCite}{rgb}{0,.7,0}
\definecolor{colorURL}{rgb}{0,0,0.7}
\usepackage[colorlinks=true,linktocpage=true,linkcolor=colorLink,citecolor=colorCite,urlcolor=colorURL]{hyperref}

\newcommand{\pt}{\ensuremath{p_T}}
\newcommand{\HT}{\ensuremath{H_T}}
\newcommand{\meff}{\ensuremath{m_\text{eff}}}

\newcommand{\NegSpace}{\vspace{-10pt}}

\newcommand{\OO}{\ensuremath{\mathcal{O}}}

\def\be{\begin{equation}}
\def\ee{\end{equation}}
\newcommand{\beq}{\begin{equation}}
\newcommand{\eeq}{\end{equation}}
\def\bea{\begin{eqnarray}}
\def\eea{\end{eqnarray}}

\newcommand{\GeV}{{\text{ GeV}}}
\newcommand{\TeV}{{\text{ TeV}}}

\newcommand{\MET}{\ensuremath{E_{T}^{\mathrm{miss}}}}

\def\ack{\section*{Acknowledgments}}

\begin{document}


%
\begin{titlepage}
\title{\Large A Comparison of Future Proton Colliders \\
Using SUSY Simplified Models:\\ A Snowmass Whitepaper}
\vspace{-30pt}
 \begin{Authlist}
Timothy Cohen
\vspace{-5pt}
\Instfoot{slac}{SLAC National Accelerator Laboratory, Menlo Park, USA}
Tobias Golling
\vspace{-5pt}
\Instfoot{yale}{Yale University, New Haven, USA}
Mike Hance
\vspace{-5pt}
\Instfoot{lbl}{Lawrence Berkeley National Laboratory, Berkeley, USA}
Anna Henrichs
\vspace{-5pt}
\Instfoot{yale}{Yale University, New Haven, USA}
Kiel Howe
\vspace{-5pt}
\Instfoot{sitp}{Stanford Institute for Theoretical Physics, Stanford University, Stanford, USA}
Joshua Loyal
\vspace{-5pt}
\Instfoot{yale}{ Yale University, New Haven, USA}
Sanjay Padhi
\vspace{-5pt}
 \Instfoot{ucsd}{University of California, San Diego, USA}
Jay G. Wacker
\vspace{-5pt}
\Instfoot{slac}{SLAC National Accelerator Laboratory, Menlo Park, USA}
\end{Authlist}

\begin{abstract}
We present a summary of results for SUSY Simplified Model searches at future proton colliders: the 14 TeV LHC as well as a 33 TeV proton collider and a 100 TeV proton collider.  Upper limits and discovery significances are provided for the gluino-neutralino (for both light and heavy flavor decays), squark-neutralino, and gluino-squark Simplified Model planes.  Events are processed with the Snowmass combined detector and Standard Model backgrounds are computed using the Snowmass samples.   We place emphasis on comparisons between different collider scenarios, along with the lessons learned regarding the impact of systematic errors and pileup.  More details are provided in a companion paper.
\end{abstract}
\end{titlepage}

\clearpage
\setcounter{page}{2}
\setstretch{1.1}
\section{Introduction}
In many proposals for physics beyond the Standard Model (BSM), the mass scale of new colored particles is tied to the degree of fine-tuning in the Higgs sector and/or the presence of a weak-scale thermal dark matter candidate. At hadron colliders, the production of these colored states often provides the first signature of new physics. The discovery and exclusion reach for new colored particles is an important metric for evaluating the physics capabilities of energy and luminosity upgrades for the LHC.  

The purpose of the Snowmass process is to evaluate the physics case for future experiments.  To that end, this document provides a series of comparisons between the following proposed proton collider scenarios:

\vspace{-10pt}
\begin{center}
\renewcommand{\arraystretch}{1.5}
\setlength{\tabcolsep}{12pt}
\begin{tabular}{c|c|c}
Machine & $\sqrt{s}$ & Final Integrated Luminosity\\
\hline
LHC Phase I & $14 \,\TeV$ & $300$ fb$^{-1}$ \\
HL-LHC or LHC Phase II &  $14 \,\TeV$ & $3000$ fb$^{-1}$ \\
HE-LHC & $33 \,\TeV$ & $3000$ fb$^{-1}$ \\
VLHC & $100 \,\TeV$ & $3000$ fb$^{-1}$
\end{tabular}
\end{center}
In order to provide quantitative results, searches must be cast in terms of the discovery reach and exclusion limits on the parameter space of specific theoretical models.  The minimal supersymmetric Standard Model (MSSM) is one of the canonical frameworks for BSM studies \cite{Martin:1997ns}. The large cross section for the production of gluinos and first and second generation squarks can yield the dominant collider signal of superpartners.  It has often been assumed that supersymmetry (SUSY) will first appear in ``jets plus missing energy" $\big(\MET\big)$ searches.  

There are a variety of SUSY paradigms worth considering.  For example, models of ``Mini-split SUSY" have received significant attention since the measurement of the Higgs boson mass at $125$ GeV~\cite{Wells:2004di, ArkaniHamed:2004fb, Giudice:2004tc, Arvanitaki:2012ps, ArkaniHamed:2012gw}.  Another well motivated scenario has been dubbed ``natural SUSY," where the light-flavor squarks are decoupled \cite{Dimopoulos:1995mi, Cohen:1996vb}.  In these models, searches for gluinos whose decays involve third generation squarks are relevant.  

While this theoretical guidance is useful for prioritizing our interest in certain regions of the MSSM, the collider signatures of colored sparticles depend heavily on the details of the superpartner spectrum.  In order to avoid obscuring the physics of the searches presented below, we take a signature-driven approach when choosing which models to study.  Generically, $R$-parity conserving SUSY theories lead to some combination of jets, \MET, and heavy flavor decays; the minimal particle content which leads to a given final state is isolated.  These ``Simplified Models" have the advantage that they tend to contain a minimal number of parameters.  Thorough explorations of their phenomenology over a wide kinematic range are possible \cite{Alwall:2008va, Alwall:2008ag, Alves:2011wf}.  

Based on this balance of theory intuition with a desire to cover a range of signatures, we choose to study the following Simplified Models: 

\vspace{-10pt}
\renewcommand{\arraystretch}{1.4}
\setlength{\tabcolsep}{8pt}
\begin{center}
\begin{tabular}{c|c|c}
Section & Simplified Model & Decay Channel\\
\hline
\ref{sec:GNLightFlavor} & Gluino-neutralino with light flavor decays &  $\widetilde{g} \rightarrow q\,\overline{q}\,\widetilde{\chi}_1^0$ \\
\ref{sec:QN} &  Squark-neutralino & $\widetilde{q} \rightarrow q\,\widetilde{\chi}_1^0$ \\
\ref{sec:GluinoSquarkNeutralino}& Gluino-squark with a massless neutralino & $\widetilde{g} \rightarrow \big (q\,\overline{q}\,\widetilde{\chi}_1^0 / q\, \widetilde{q}^*\big)$; $\widetilde{q} \rightarrow \big( q\,\widetilde{\chi}_1^0 / q\,\widetilde{g} \big) $ \\
 \ref{sec:GNHeavyFlavor}& Gluino-neutralino with heavy flavor decays &  $\widetilde{g} \rightarrow t\,\overline{t}\,\widetilde{\chi}_1^0$ 
\end{tabular}
\end{center}
This set of models, which is by no means exhaustive, covers a wide range of the expected signatures from SUSY. These models also provide a good stand-in for a variety of non-SUSY models where similar final states are important.

A particular final state informs the choice of search strategy.  Furthermore, the compressed regions of parameter space, where two masses become degenerate, also requires a targeted approach.  Since the parameter space of SUSY Simplified Models has been explored in great detail at the 8 TeV LHC by both the ATLAS and CMS collaborations, we will tend to follow existing public search strategies with optimizations performed to account for the higher luminosity and energy.  

Our studies rely on the Snowmass backgrounds \cite{Avetisyan:2013onh} and detector framework \cite{Anderson:2013kxz} and OSG grid \cite{Avetisyan:2013dta}.  The paper containing the details of our studies \cite{Cohen:2013xda} also includes some validation plots which demonstrates good agreement with a 14 TeV ATLAS study \cite{ATL-PHYS-PUB-2013-002}. Together these results provide a coherent picture of the expected performance of the LHC upgrades, including the sensitivity to assumptions about pile-up and systematic uncertainties.  The purpose of this note is to provide a succinct summary of the results that are relevant for the Snowmass process.

The rest of this paper is organized as follows.  We first present the search strategies used to confront the simplified models described above.  Each of the following four sections are then devoted to a different Simplified Model, including both theoretical details and the results of the studies using one or more search strategies.  A discussion of lessons learned from the detailed study is provided in Sec.~\ref{sec:Lessons}.  Full descriptions of our techniques and results are given in a companion paper \cite{Cohen:2013xda}.

\section{Analysis Strategies}
\label{sec:Analyses}

In the studies presented here, we use four different analysis strategies in order to probe the full range of collider scenarios and Simplified Models.  Each strategy defines a signature and one or more discriminating variables, and then optimizes cuts on those variables based on each collider scenario and signal hypothesis.  

\subsection{Jets + $\boldmath{E_T}^{\!\!\!\!\text{\bf{miss}}}$}

The strategy we refer to as ``jets + \MET'' is similar to the ATLAS studies presented in~\cite{ATL-PHYS-PUB-2013-002}.  For all collider scenarios and Simplified Models, we pre-select events by requiring $\MET>100$ GeV and the presence of at least four jets with $\pt>60$ GeV.   We further require $\MET/\sqrt{\HT}>15\GeV^{1/2}$, which suppresses backgrounds from Standard Model multijet production to negligible levels~\cite{ATL-PHYS-PUB-2013-002}.  We veto events with electrons or muons to suppress backgrounds from $W$+jets.  In order to reduce contributions from $W$ or $Z$ bosons produced in association with a single hard jet, we require that the leading jet carry less than 40\% of the total \HT.  Finally, cuts on \MET\ and \HT\ are simultaneously optimized for each simulated point in parameter space.

\subsection{Compressed Spectrum}
The ``compressed spectrum'' search strategy is employed in scenarios where the mass spectrum has a small gap between the superpartner produced in the collisions and the neutralino:
\be
m_{\widetilde{g,q}} - m_{\widetilde{\chi}^0} \equiv \Delta m \ll m_{\widetilde{g,q}}.
\ee
The partons that result from the sparticle decay are soft, leading to signatures with reduced \HT~and \MET.  Therefore it is useful to rely on events with hard jets from initial state radiation to discriminate signal from background.  We construct four signal regions optimized for this kinematic configuration and choose the one that leads to the most stringent bound on the production cross section for each point in parameter space.  

All compressed-spectrum searches have a common preselection.  We require events to have $\MET > 100$ GeV, and to have a central leading jet with $\pt>30$ GeV and $|\eta|<2.5$. Events with electrons or muons are vetoed, as in the jets + \MET~analysis.  We then define two signal regions:

\begin{enumerate}
\item Events are selected with at most two jets with $\pt>30$ GeV, as long as the second jet and \MET~ are well-separated in $\phi$.  A simultaneous optimization is performed over a $p_T$ cut on the leading jet and a cut on \MET.  
\item No requirement on the number of jets is made.  An optimization over a cut on \MET~is preformed.  
\end{enumerate}

Our results use the most sensitive signal region for each collider scenario and signal hypothesis.

\subsection{Same-sign Dilepton}

For models that produce multiple prompt leptons, we define a same-sign dilepton search strategy.  We select events with pairs of same-sign electrons or muons with at least two $b$-tagged jets.  A veto is applied for events where a third lepton combines with one of the same-sign leptons to form a mass within 15 GeV of the $Z$-boson mass.  Eight signal regions are defined using a wide variety of variables: $M_{T2}$, the $p_T$ of the hardest lepton, \MET, \HT, \meff, and the number of jets with and without $b$-tags.  Each choice of cuts is optimized for a particular region of the Simplified Model parameter space.  Some signal regions specifically target compressed spectra by relaxing cuts on \MET.

Now that the three analysis strategies have been outlined, we can discuss their applications to a variety of Simplified Models.

\section{The Gluino-Neutralino Model with Light Flavor Decays}
\label{sec:GNLightFlavor}
In the ``gluino-neutralino model with light flavor decays", the gluino is the only kinematically accessible colored particle. The squarks are completely decoupled and do not contribute to gluino production diagrams. The gluino undergoes a prompt three-body decay through off-shell squarks,  $\widetilde{g} \rightarrow q\,\overline{q}\,\widetilde{\chi}^0$, where $q = u,d,c,s$ are the first and second generation quarks and $\widetilde{\chi}_1^0$ is a neutralino. The branching ratios to different light quarks are taken to be equal. The only two relevant parameters are the gluino mass $m_{\widetilde{g}}$ and the neutralino mass $m_{\widetilde{\chi}_1^0}$.  This model can be summarized by:
\begin{center}
\NegSpace
\renewcommand{\arraystretch}{1.3}
\setlength{\tabcolsep}{12pt}
\begin{tabular}{c|c|c}
BSM particles & production & decays \\
\hline
$\widetilde{g},\,\widetilde{\chi}_1^0$ & $p\,p\rightarrow \widetilde{g}\, \widetilde{g}$ & $\widetilde{g} \rightarrow q\,\overline{q}\,\widetilde{\chi}_1^0 $ 
\end{tabular}
\end{center}

One motivation comes from (mini-)split supersymmetry scenarios~\cite{Wells:2004di, ArkaniHamed:2004fb, Giudice:2004tc, Arvanitaki:2012ps, ArkaniHamed:2012gw}, where the scalar superpartners are heavier than the gauginos.  The final state is four (or more) hard jets and missing energy, and the most important backgrounds are $W/Z+\text{jets}$ (which dominates at $\sqrt{s}=14\TeV$) and $t\overline{t}$ (which dominates at $\sqrt{s}=100\TeV$).  This Simplified Model provides a good proxy for comparing the power of searches which rely on jets+$\MET$ search strategies.

\begin{figure}[h!]
  \centering
  \includegraphics[width=.48\columnwidth]{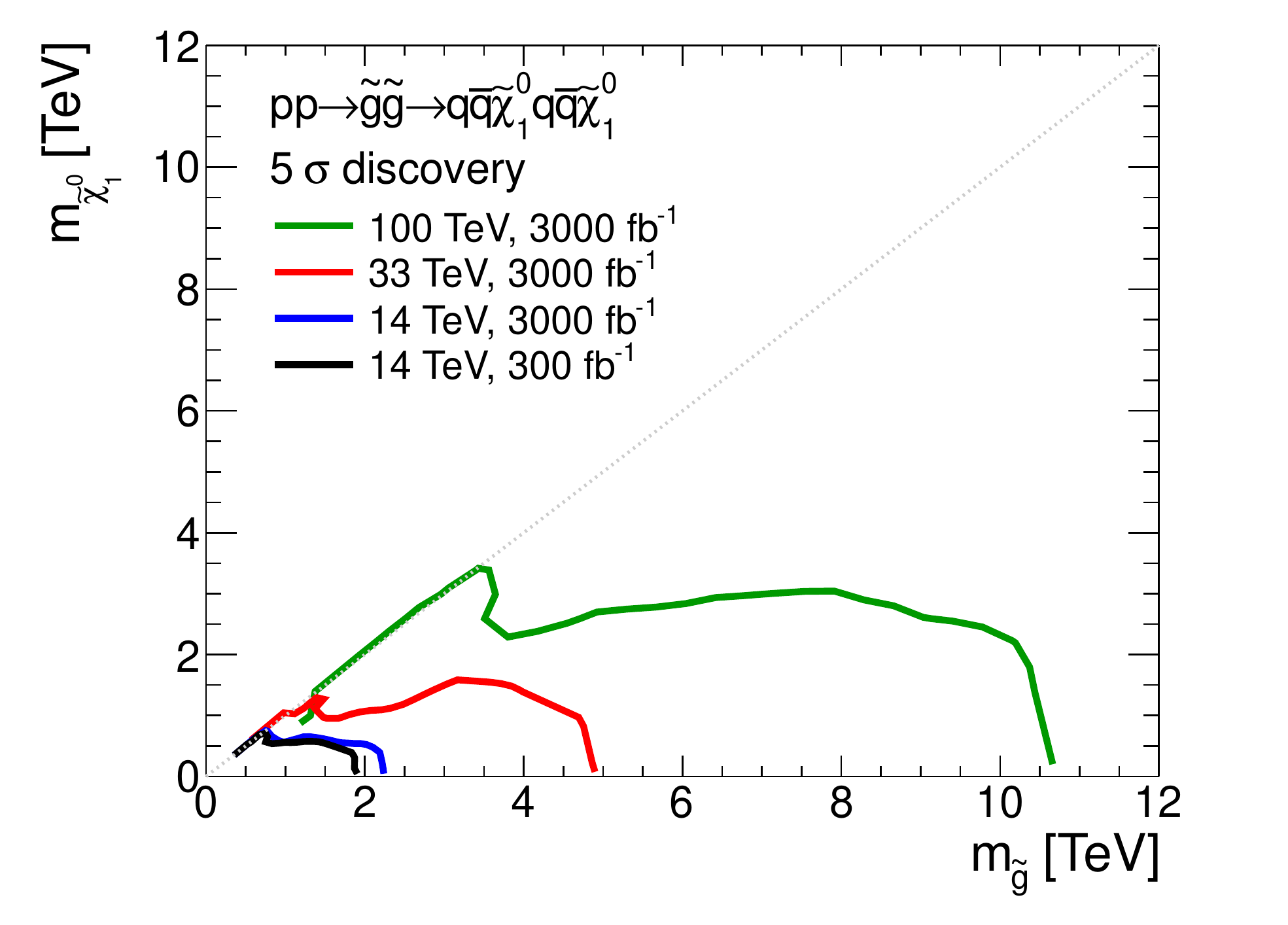}
  \includegraphics[width=.48\columnwidth]{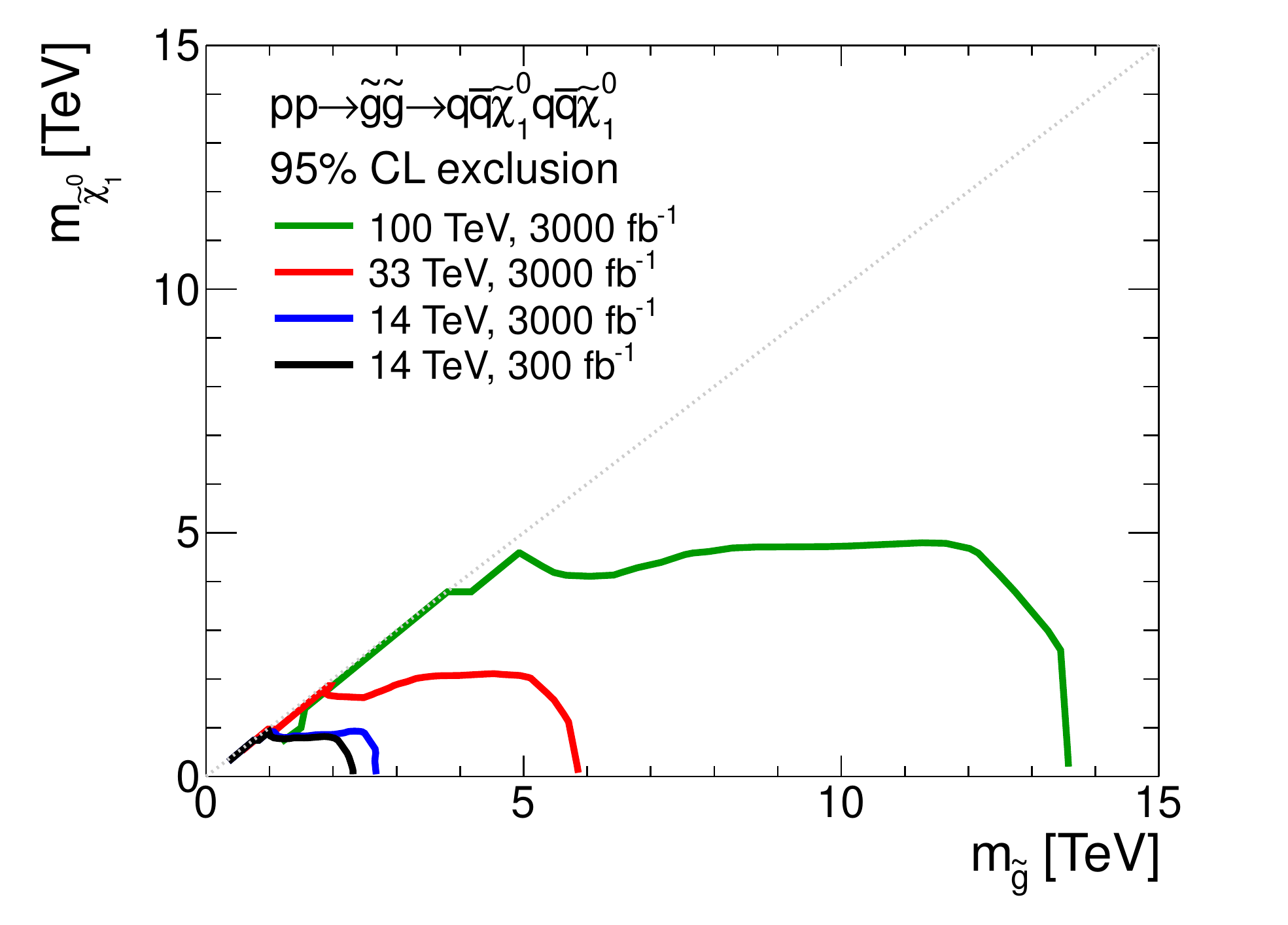}
  \caption{Results for the gluino-neutralino model with light flavor decays using the jets + \MET~analysis strategy.  The left [right] panel shows the $5\,\sigma$ discovery reach [$95\%$ CL exclusion] for the four Snowmass collider scenarios.  A 20\% systematic error is assumed and pileup is not included.}
    \label{fig:GOGO_Comparison}
\end{figure}

Results for the gluino-neutralino model with light flavor decays using the jets + \MET\ search are shown in Fig.~\ref{fig:GOGO_Comparison}.  The left [right] plot gives the $5\,\sigma$ discovery reach [$95\%$ CL exclusion].  For a massless neutralino, the $14 \TeV$ LHC could discover a $\sim 2 \TeV$ gluino, a $33 \TeV$ proton collider could discover a $\sim 5 \TeV$ gluino, and a $100 \TeV$ proton collider could discover a $\sim 11 \TeV$ gluino.  This search is also sensitive to the region where $m_{\widetilde{g}} \simeq m_{\widetilde{\chi}^0}$ and could lead to a discovery of a model with $m_{\widetilde{g}} \simeq m_{\widetilde{\chi}_1^0} \sim 3.5 \TeV$ at a 100 TeV machine.

\begin{figure}[!h]
\begin{center}
\includegraphics[width=0.48\textwidth]{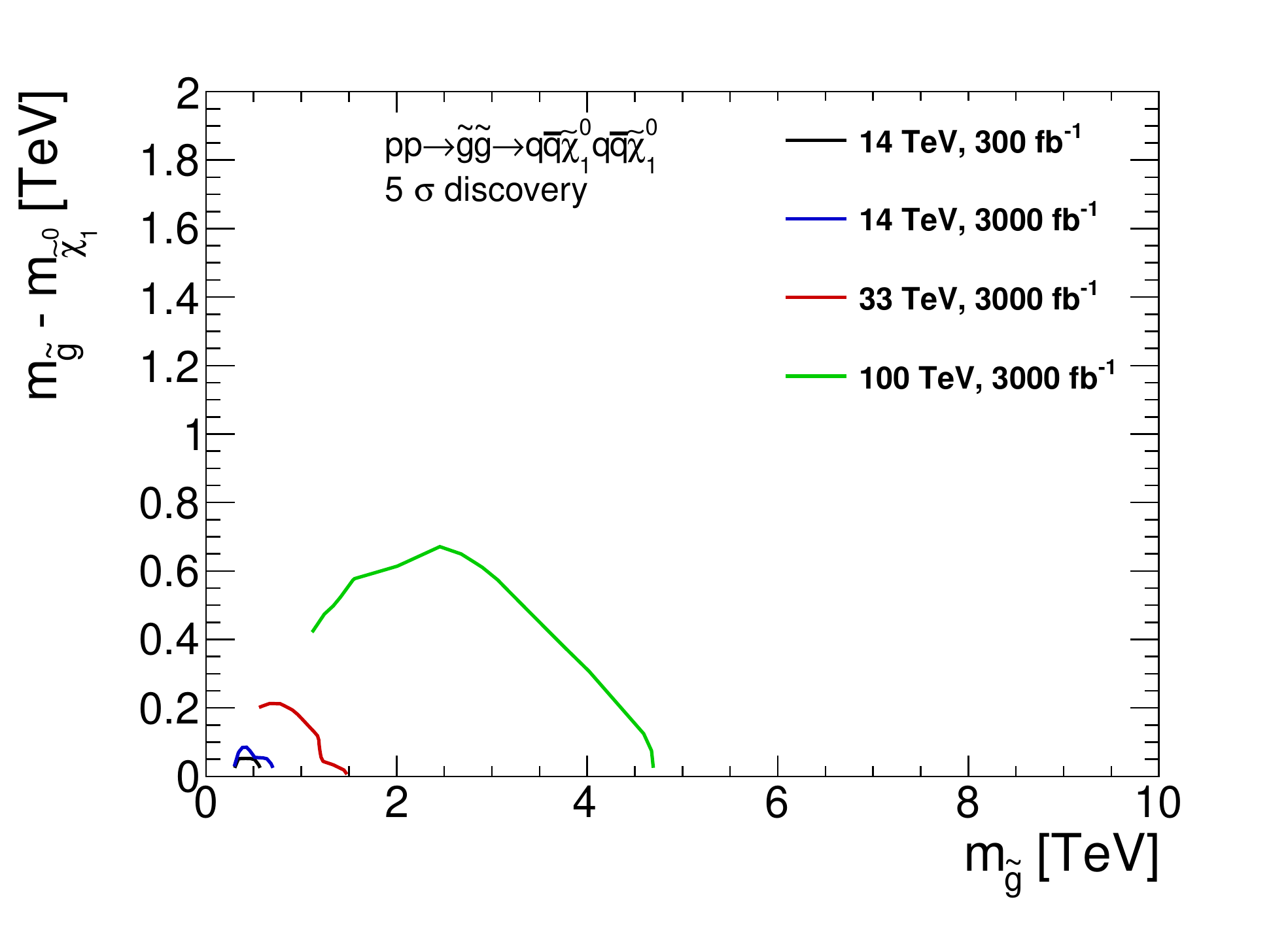}
\includegraphics[width=0.48\textwidth]{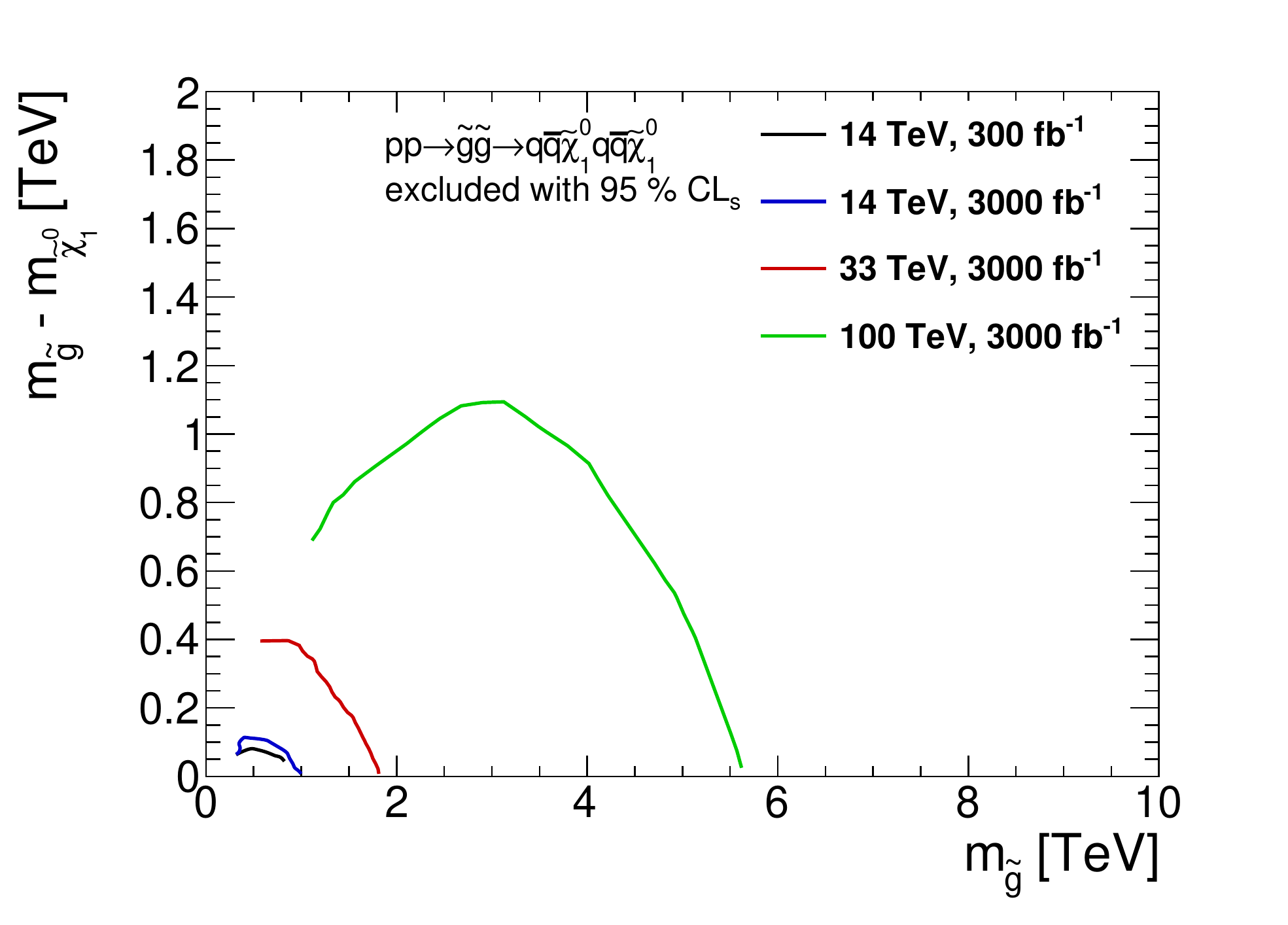}
\caption{Results for the compressed region of the gluino-neutralino model with light flavor decays using the compressed spectrum analysis strategy.  The left [right] panel shows the $5\,\sigma$ discovery reach [$95\%$ CL exclusion] for the four Snowmass collider scenarios.  A 20\% systematic error is assumed and pileup is not included.}
\label{fig:GNLightFlavorCompressedResults}
\end{center}
\end{figure}

Results for the same model using the compressed-spectrum search are shown in Fig.~\ref{fig:GNLightFlavorCompressedResults}.  The left [right] plot gives the $5\,\sigma$ discovery reach [$95\%$ CL exclusion].  For $m_{\widetilde{g}} \simeq m_{\widetilde{\chi}^0}$, the $14 \TeV$ HL-LHC could discover a $\sim 700 \GeV$ gluino, a $33 \TeV$ proton collider could discover a $\sim 1.5 \TeV$, and a $100 \TeV$ proton collider could discover a $\sim 4.6 \TeV$ gluino.  Note that the 100 TeV result leads to a $\sim30\%$ improvement when utilizing searches that target the compressed spectra directly when compared to the results for the bulk of the parameter space.

\pagebreak

\section{The Squark-Neutralino Model}
\label{sec:QN}
In the ``squark-neutralino model", the first and second generation squarks
 $\widetilde{q} = \widetilde{u}_L, \widetilde{u}_R$, $\widetilde{d}_L, \widetilde{d}_R,$ $\widetilde{c}_L, \widetilde{c}_R,$ $\widetilde{s}_L, \widetilde{s}_R$ 
are the only kinematically accessible colored states. The gluino is completely decoupled from the squark production diagrams --- the squark production cross section is significantly reduced when compared to models where the gluino is just above the kinematic limit (see Sec.~\ref{sec:GluinoSquarkNeutralino} below). The squarks decay directly to the LSP and the corresponding quark, $\widetilde{q}_i \rightarrow q_i\,\widetilde{\chi}_1^0$. The only two relevant parameters are the squark mass $m_{\widetilde{q}}$, which is taken to be universal for the first two generations, and the neutralino mass $m_{\widetilde{\chi}_1^0}$.   The model is summarized as:
\begin{center}
\NegSpace
\renewcommand{\arraystretch}{1.3}
\setlength{\tabcolsep}{12pt}
\begin{tabular}{c|c|c}
BSM particles & production & decays \\
\hline
$\widetilde{q},\,\widetilde{\chi}_1^0$ & $p\,p \rightarrow \widetilde{q} \, \widetilde{q}^*$ & $\widetilde{q} \rightarrow q\,\widetilde{\chi}_1^0 $ 
\end{tabular}
\end{center}

Due to the structure of the renormalization group equations in the MSSM, a heavy gluino would tend to raise the squark masses; in order to have light squarks some tuning is required.  However, a class of theories with Dirac gluinos can be well approximated by this Simplified Model~\cite{Kribs:2013oda}. 

Since the final state is two (or more) hard jets and missing energy, this model also serves to test the power of jets + \MET~style analyses.  As with the gluino search, the dominant backgrounds are from $W/Z$+jet production and $t\,\overline{t}$.   The mass reach is not as strong as in the gluino-neutralino model with light flavor decays for two reasons: the final state has only two hard jets from the squark decays as opposed to four hard jets from the gluino decays, and cross section for producing squark pairs with the gluino completely decoupled is substantially lower than that for producing gluino pairs of the same mass. 

\begin{figure}[h!]
  \centering
  \includegraphics[width=.48\columnwidth]{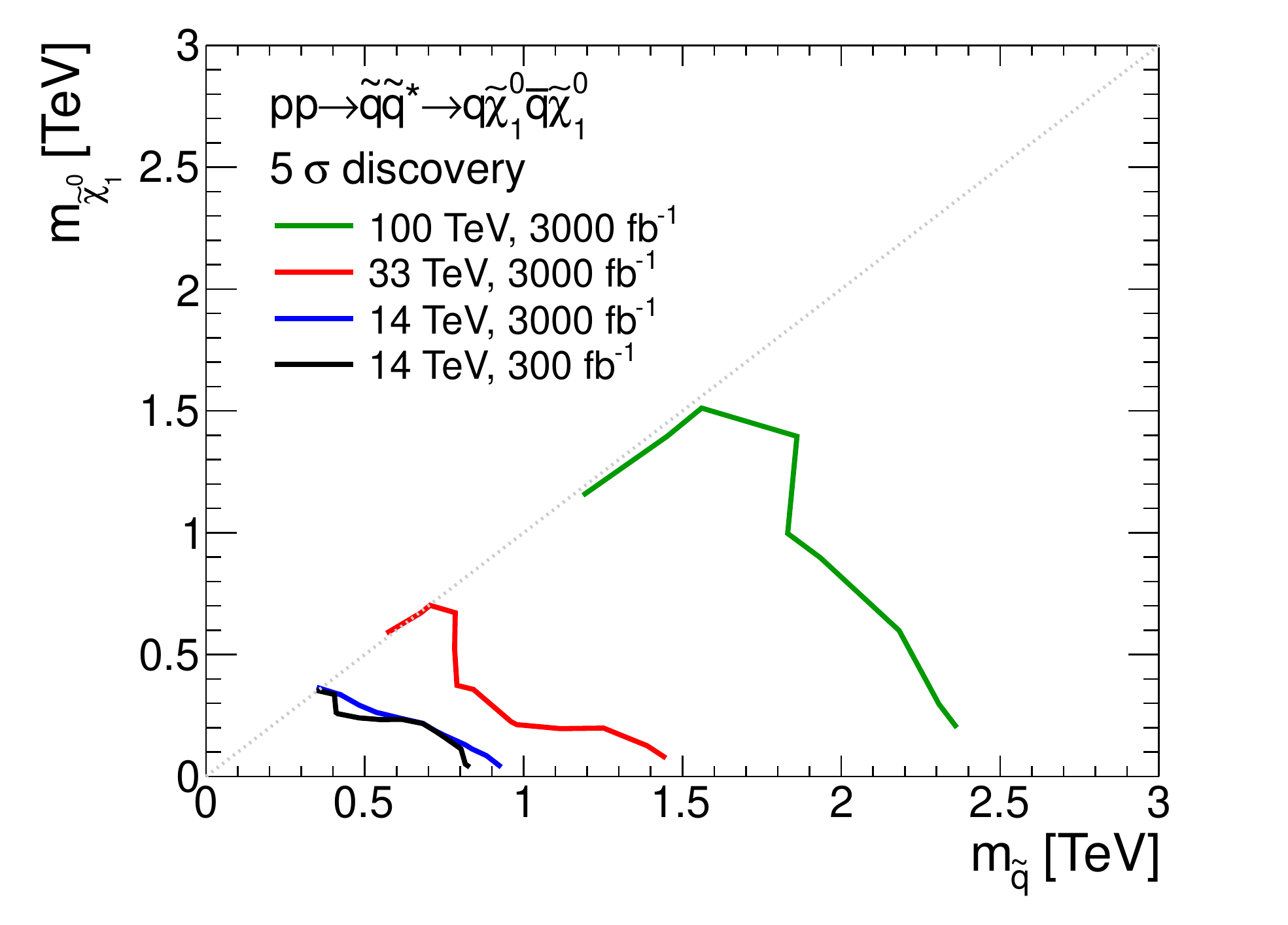}
  \includegraphics[width=.48\columnwidth]{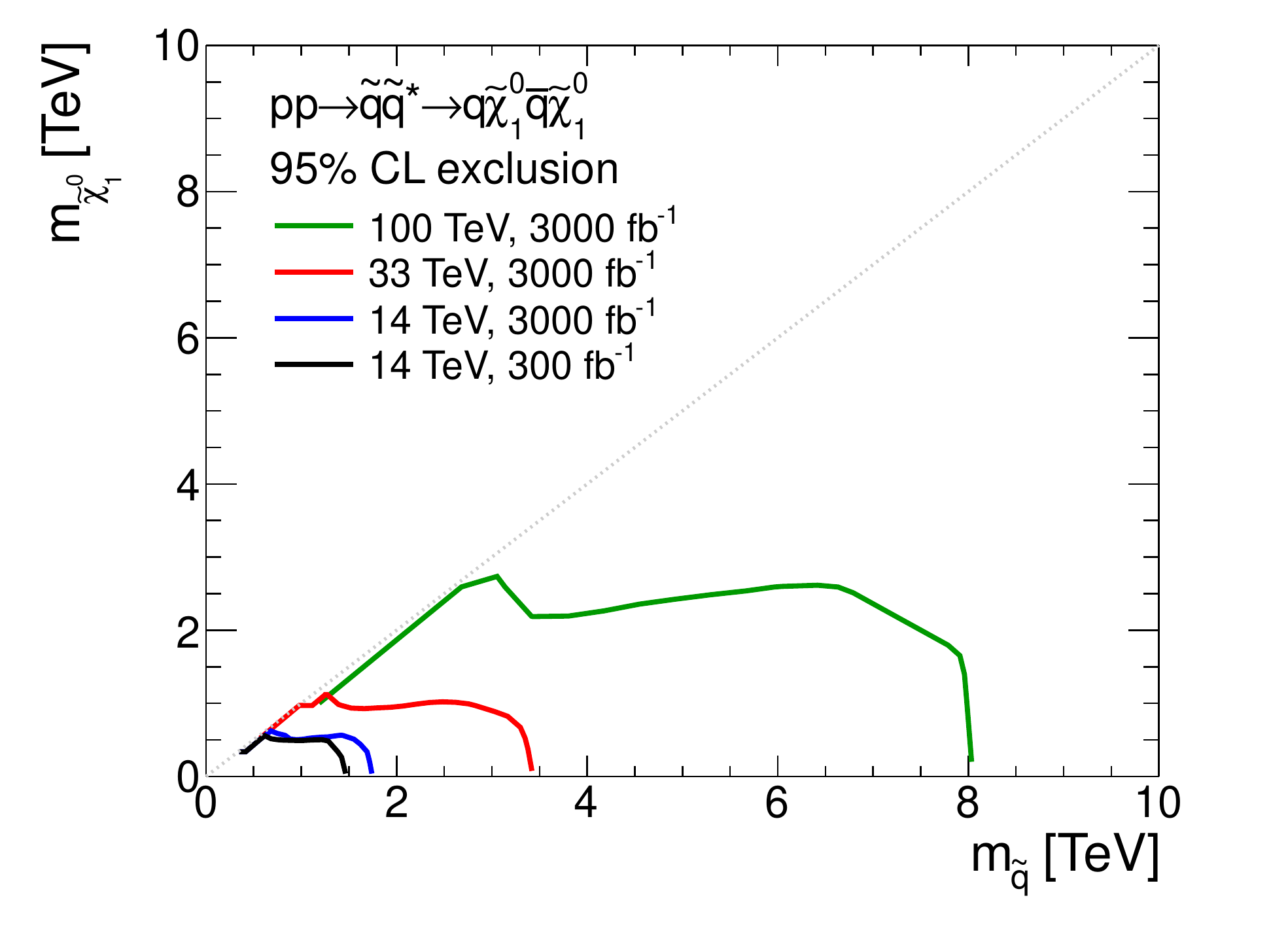}
  \caption{Results for the squark-neutralino model using the jets + \MET~analysis strategy.  The left [right] panel shows the $5\,\sigma$ discovery reach [$95\%$ CL exclusion] for the four Snowmass collider scenarios.  A 20\% systematic error is assumed and pileup is not included.}
    \label{fig:SquarkNeutralino_Comparison}
\end{figure}

Results for the squark-neutralino model using the jets + \MET\ strategy are shown in Fig.~\ref{fig:SquarkNeutralino_Comparison}.  The left [right] plot gives the $5\,\sigma$ discovery reach [$95\%$ CL exclusion].  For a massless neutralino, the $14 \TeV$ LHC could discover $\sim 900 \GeV$ squarks, a $33 \TeV$ proton collider could discover $\sim 1.4 \TeV$ squarks, and a $100 \TeV$ proton collider could discover $\sim 2.4 \TeV$ squarks.  This search is also sensitive to the region where $m_{\widetilde{q}} \simeq m_{\widetilde{\chi}^0}$ and could lead to a discovery of a model with $\sim 1.6 \TeV$ squarks and neutralino at a $100$ TeV machine.

\begin{figure}[h!]
  \centering
  \includegraphics[width=.48\columnwidth]{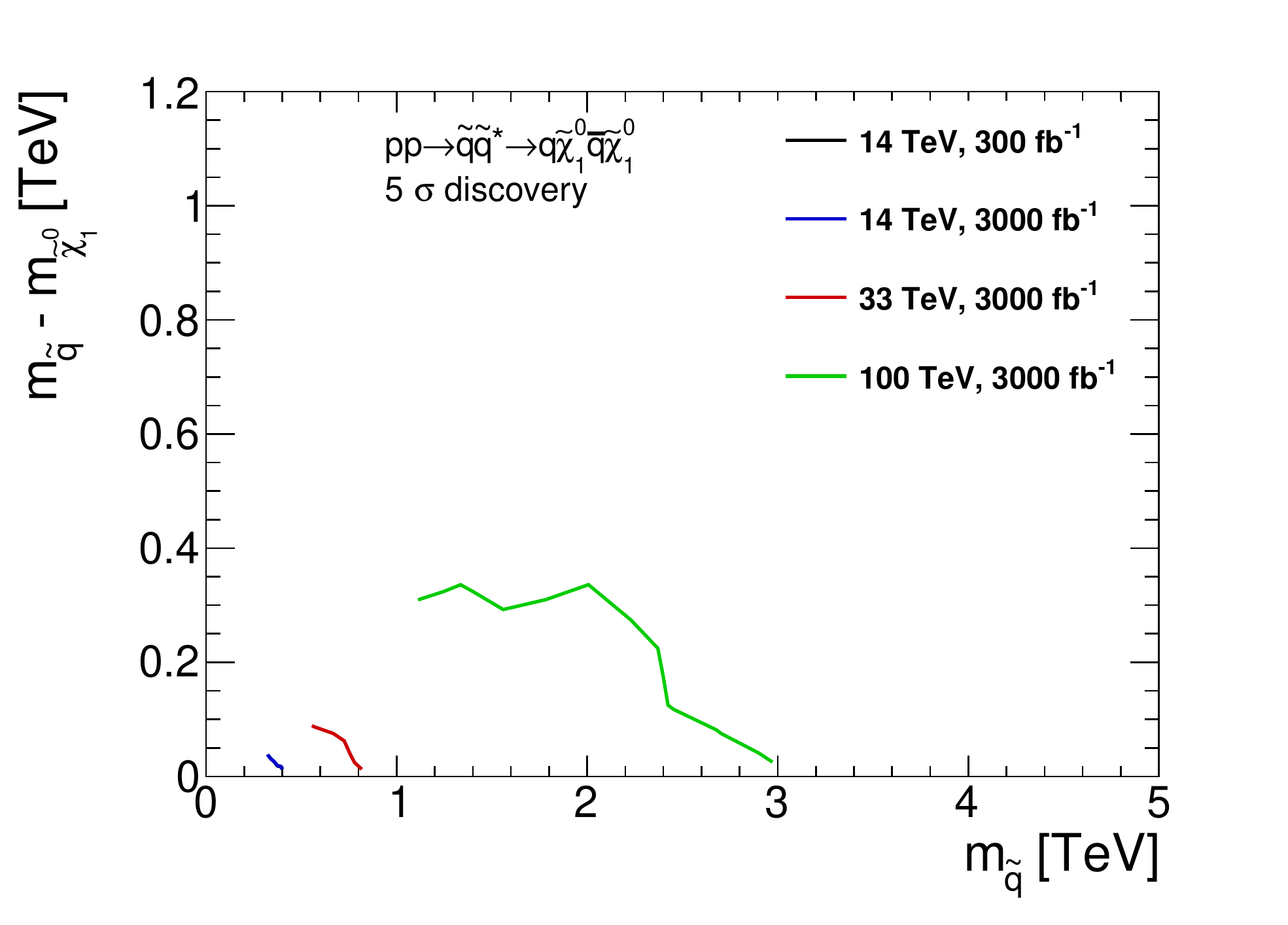}
  \includegraphics[width=.48\columnwidth]{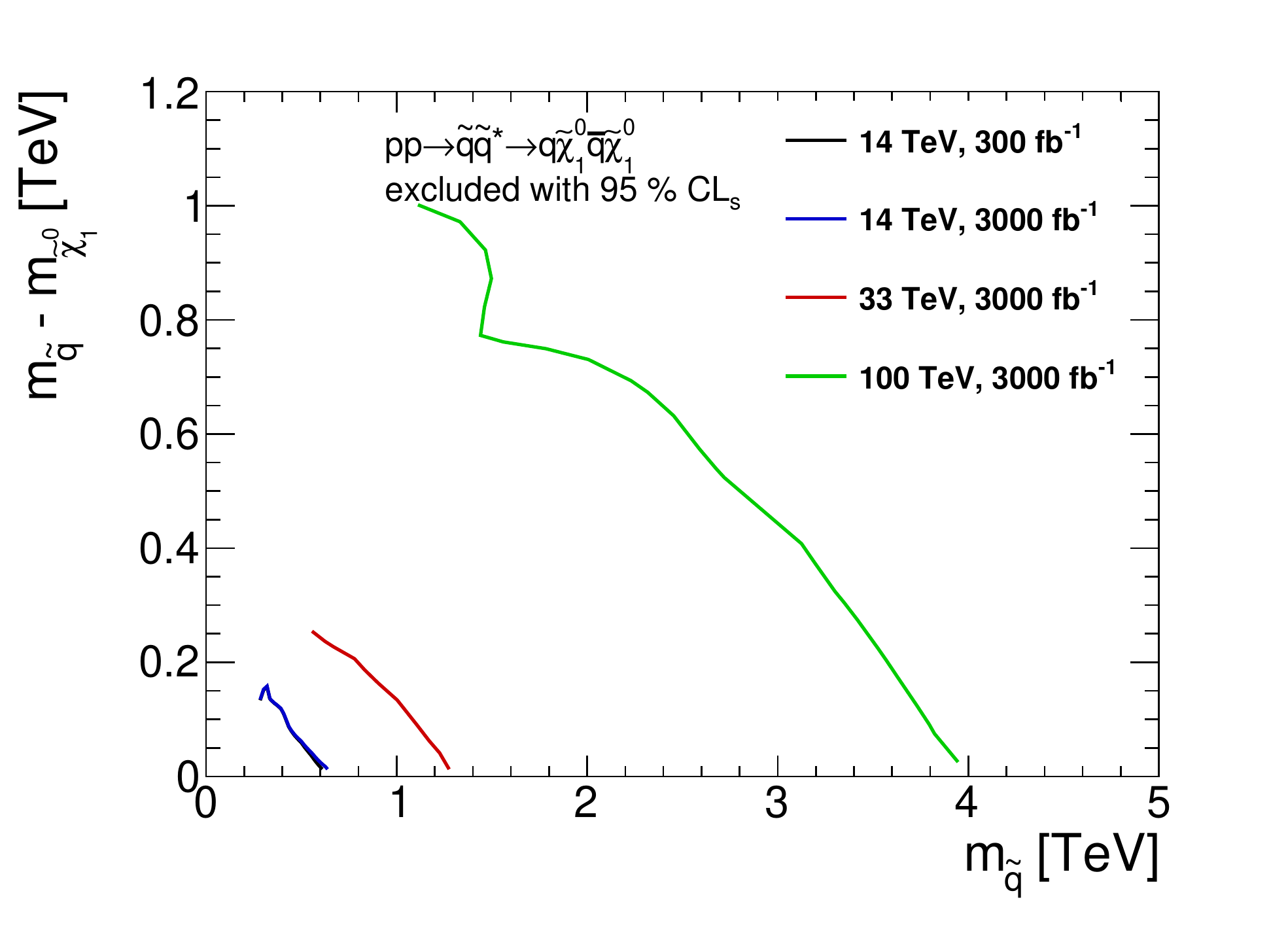}
  \caption{Results for the compressed region of the squark-neutralino model using the compressed spectrum analysis strategy.  The left [right] panel shows the $5\,\sigma$ discovery reach [$95\%$ CL exclusion] for the four Snowmass collider scenarios.  A 20\% systematic error is assumed and pileup is not included.}
    \label{fig:SquarkNeutralino_Compressed_Comparison}
\end{figure}

Results for the same model using the compressed-spectrum strategy are shown in Fig.~\ref{fig:SquarkNeutralino_Compressed_Comparison}.  The left [right] plot gives the $5\,\sigma$ discovery reach [$95\%$ CL exclusion].  For $m_{\widetilde{q}} \simeq m_{\widetilde{\chi}^0}$, the $14 \TeV$ HL-LHC could discover $\sim 500 \GeV$ squarks, a $33 \TeV$ proton collider could discover $\sim 800 \GeV$ squarks, and a $100 \TeV$ proton collider could discover $\sim 2.5 \TeV$ squarks.  Note that the 100 TeV result shows a factor of $\sim 50\%$ improvement from utilizing searches that target the compressed spectra directly as compared to the strategy relevant for the bulk of the parameter space.

\section{The Gluino-Squark-Neutralino Model}
\label{sec:GluinoSquarkNeutralino}
In the ``gluino-squark-neutralino model", the gluino and the first and second generation squarks are all allowed to be kinematically accessible. The only relevant parameters are the squark mass $m_{\widetilde{q}}$, which is taken to be universal for the first two generations, the gluino mass $m_{\widetilde{g}}$, and the neutralino mass $m_{\widetilde{\chi}^0}$. For this study we fix the neutralino mass $m_{\widetilde{\chi}^0}=1{\rm~GeV}$, which captures the relevant kinematics for $m_{\widetilde{g}},m_{\widetilde{q}}\gg m_{\widetilde{\chi}^0}$. The decay mode is chosen depending on the mass hierarchy.  This model is summarized as:
\begin{center}
\NegSpace
\renewcommand{\arraystretch}{1.6}
\setlength{\tabcolsep}{12pt}
\begin{tabular}{c|c|c}
BSM particles & production & decay \\
\hline
\rule{0pt}{5ex}
\multirow{6}{*}{$\widetilde{g},\,\widetilde{q},\,\widetilde{\chi}^0$} & $p\,p \rightarrow \widetilde{g}\,\,\widetilde{g}$ & \multirow{3}{*}{$\widetilde{g} \rightarrow \begin{cases}
\widetilde{q}\,\,\overline{q} & \text{for}\,\,\, m_{\widetilde{g}} > m_{\widetilde{q}} \\
q\,\overline{q}\,\widetilde{\chi}^0 & \text{for}\,\,\, m_{\widetilde{g}} \simeq m_{\widetilde{q}} \\
q\,\overline{q}\,\widetilde{\chi}^0 & \text{for}\,\,\,  m_{\widetilde{g}} < m_{\widetilde{q}}
\end{cases}
$} \\
 & $p\,p \rightarrow \widetilde{g}\,\,\widetilde{q}$ & \\
 & $p\,p \rightarrow \widetilde{g}\,\,\widetilde{q}^*$ & \\
 & $p\,p \rightarrow \widetilde{q}\,\,\widetilde{q}^*$ & \multirow{3}{*}{$\widetilde{q} \rightarrow \begin{cases}
q\,\widetilde{\chi}^0  & \text{for}\,\,\, m_{\widetilde{g}} > m_{\widetilde{q}} \\
q\,\widetilde{\chi}^0 & \text{for}\,\,\, m_{\widetilde{g}} \simeq m_{\widetilde{q}} \\
q\,\widetilde{g}& \text{for}\,\,\,  m_{\widetilde{g}} < m_{\widetilde{q}}
\end{cases}
$} \\
  & $p\,p \rightarrow \widetilde{q}\,\,\widetilde{q}$ &\\
  & & 
\end{tabular}
\end{center}

For a full MSSM model, which in particular would imply a specific neutralino composition, there can be a non-zero branching ratio for the squark to decay to a neutralino and a quark.  In this study, it is assumed that the squark is weakly coupled to the neutralino and decays to the gluino proceed with $100\%$ branching ratio when kinematically allowed. Likewise for $m_{\widetilde{g}} > m_{\widetilde{q}}$, the branching ratio of the gluino to $3$-body versus $2$-body decays depends on the masses and coupling of the squarks to the neutralino; we take the $2$-body branching ratio to be $100\%$ in this region of parameter space.  To capture the transition region where the gluino and squark are nearly degenerate, parameter choices along the line $m_{\widetilde{g}} = m_{\widetilde{q}}$ are included; the gluino decay is taken to be $3$-body and the squarks are assumed to decay directly to the neutralino.


This model is a good proxy for comparing the power of searches which rely on jets and $\MET$ to discriminate against background.  The final state ranges from two to four (or more) hard jets from the decay (depending on the production channel) and missing energy.  As with the gluino and squark models, the dominant backgrounds are from $W/Z$+jet production and $t\,\overline{t}$.   Additionally, the squark production cross section has a strong dependence on the gluino mass which impacts the reach with respect to the results in Sec.~\ref{sec:QN}.

\begin{figure}[h!]
  \centering
 \includegraphics[width=.48\columnwidth]{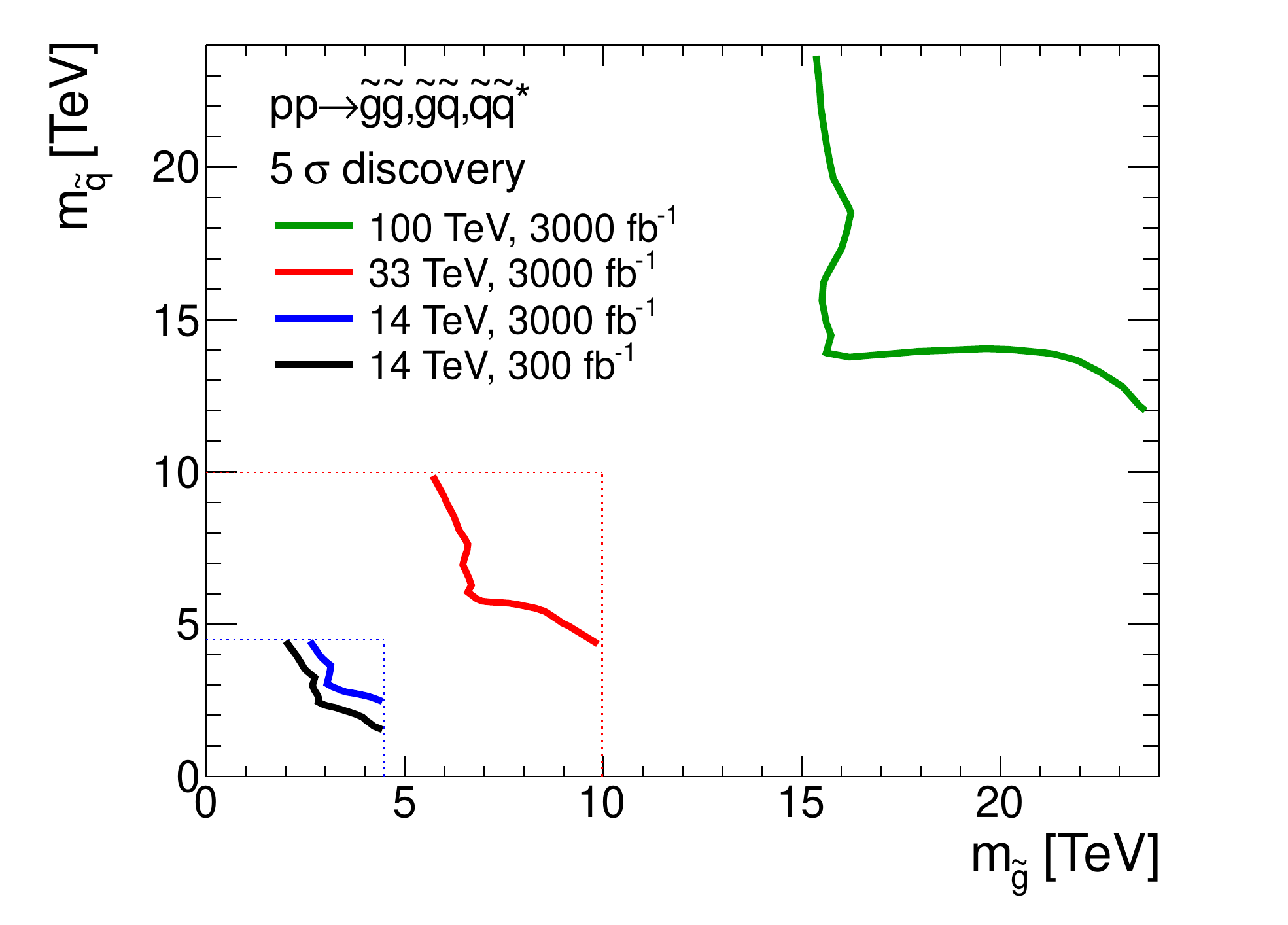}
\includegraphics[width=.48\columnwidth]{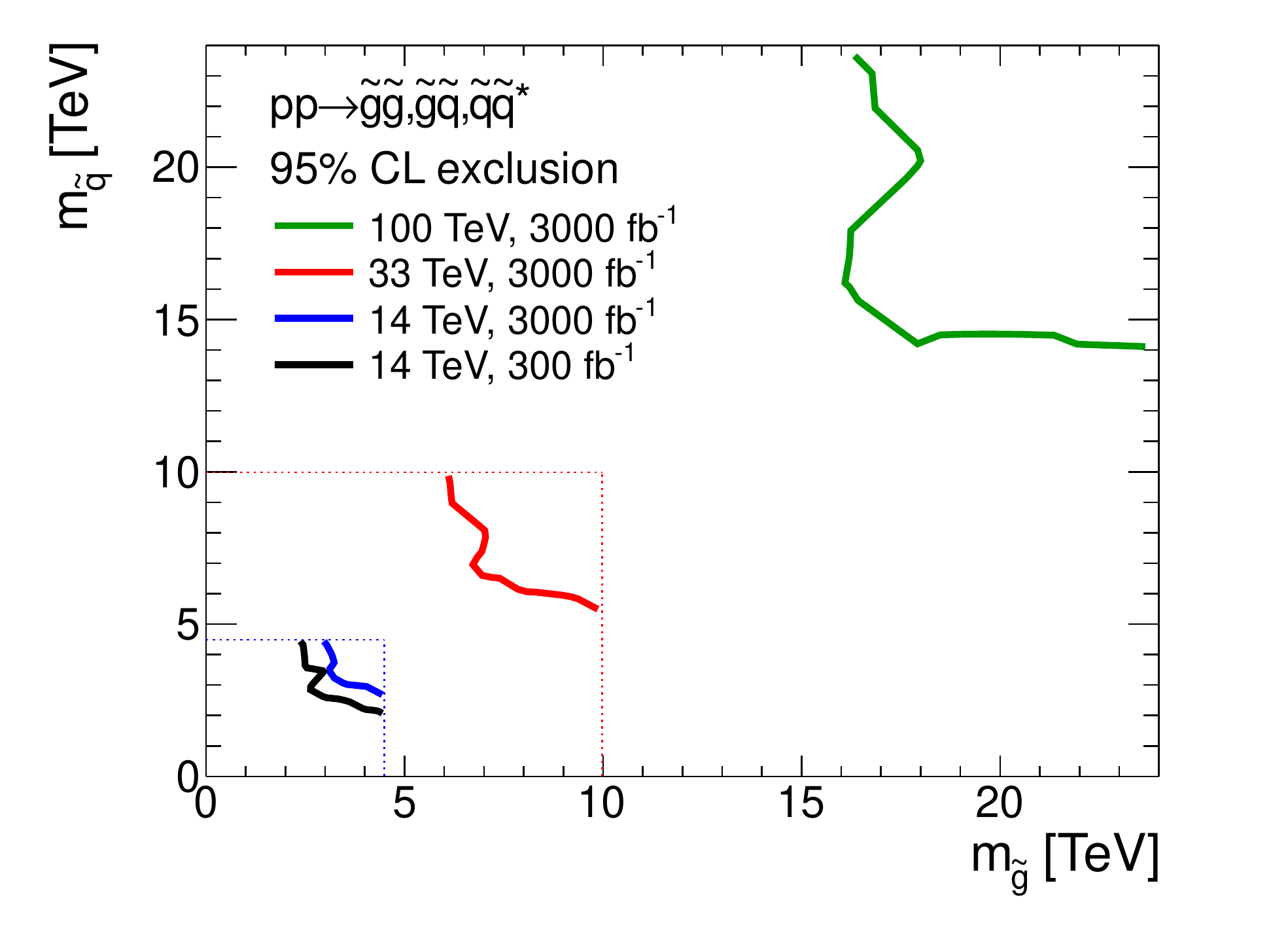}
  \caption{Results for the gluino-squark model with a 1 GeV neutralino using the jets + \MET~analysis strategy.  The left [right] panel shows the $5\,\sigma$ discovery reach [$95\%$ CL exclusion] for the four Snowmass collider scenarios.  The dotted lines the boundaries for our scans in gluino and squark mass.  A 20\% systematic error is assumed and pileup is not included.}
    \label{fig:GluinoSquark_Comparison}
\end{figure}

Results for the gluino-squark model with a 1 GeV neutralino are shown in Fig.~\ref{fig:GluinoSquark_Comparison}.  The left [right] plot gives the $5\,\sigma$ discovery reach [$95\%$ CL exclusion].  The dotted boundaries show the range of scanned gluino and squark masses for each collider scenario.  Note that the maximum masses along these decoupling directions are not large enough to approach the pure gluino-neutralino/squark-neutralino models presented above.  The $14 \TeV$ HL-LHC could discover a model with $m_{\widetilde{g}} \simeq m_{\widetilde{q}} \sim 3 \TeV$, a $33 \TeV$ proton collider could discover a model with $m_{\widetilde{g}} \sim 7 \TeV$ and $m_{\widetilde{q}} \sim 6 \TeV$, and a $100 \TeV$ proton collider could discover a model with $m_{\widetilde{g}} \sim 16 \TeV$ and $m_{\widetilde{q}} \sim 14 \TeV$.

\pagebreak
\section{The Gluino-Neutralino Model with Heavy Flavor Decays}
\label{sec:GNHeavyFlavor}
In the ``gluino-neutralino model with heavy flavor decays", the gluino $\widetilde{g}$ is the only kinematically accessible colored particle. The squarks are completely decoupled and do not contribute to gluino production diagrams. The gluino undergoes a prompt three-body decay through off-shell stops,  $\widetilde{g} \rightarrow t\,\overline{t}\,\widetilde{\chi}^0_1$, where $t$ is the top quark and $\widetilde{\chi}^0_1$ is a neutralino LSP. The only two relevant parameters are the gluino mass $m_{\widetilde{g}}$ and the neutralino mass $m_{\widetilde{\chi}^0_1}$.  This model can be summarized by:
\begin{center}
\NegSpace
\renewcommand{\arraystretch}{1.3}
\setlength{\tabcolsep}{12pt}
\begin{tabular}{c|c|c}
BSM particles & production & decays \\
\hline
$\widetilde{g},\,\widetilde{\chi}^0_1$ & $p\,p\rightarrow \widetilde{g}\, \widetilde{g}$ & $\widetilde{g} \rightarrow t\,\overline{t}\,\widetilde{\chi}^0_1 $ 
\end{tabular}
\end{center}

This model has a variety of motivations.  Perhaps the most compelling are ``natural" SUSY scenarios \cite{Dimopoulos:1995mi, Cohen:1996vb, Papucci:2011wy, Brust:2011tb, Essig:2011qg}, where the stop mass is assumed to be below the (stronger) bounds on first and second generation squark masses; for some examples of explicit constructions, see \cite{ArkaniHamed:1997fq, Craig:2012hc, Craig:2012di, Csaki:2012fh, Larsen:2012rq, Cohen:2012rm, Randall:2012dm}.  If both the stop and gluino are kinematically accessible for a given center-of-mass energy, the gluino would be visible above background before that of the stop; this Simplified Model reproduces the first signature of this paradigm.  Note that in these models, the gluino decays involving on-shell stops.  However, the final state are identical and the kinematics are similar enough that the reach is qualitatively reproduced by the results presented below.

There is a class of split-SUSY models where the inaccessible stops are somewhat lighter than the other squarks --- this Simplified Model acts as an excellent proxy for the first signatures of these scenarios.  There are compelling reasons to believe this is a ``preferred" spectrum.  Renormalization group evolution tends to reduce the stop mass with respect to the first/second generation squarks (due to the large top Yukawa coupling) \cite{Martin:1997ns}.  Also, avoiding flavor and/or CP violation bounds would imply that the squarks have masses greater than $\OO(1000 \TeV)$ \cite{Altmannshofer:2013lfa}, while achieving a Higgs boson with a mass $\sim125 \GeV$ is consistent with a stop mass less than $\OO(100 \TeV)$, assuming the MSSM \cite{Giudice:2011cg}.

The model produces two $t\,\overline{t}$ pairs along with considerable \MET~(away from the compressed region of parameter space).  For this search we use the same-sign dilepton strategy, which reduces the Standard Model backgrounds significantly.  The dominant remaining background is top pair production, where one or both tops decay leptonically and a $b$-jet decays semi-leptonically and thus misidentified as an additional lepton.  There are subdominant backgrounds from rare decays, di-boson, tri-boson, and single top production.

\begin{figure}[h!]
  \centering
\includegraphics[width=.48\columnwidth]{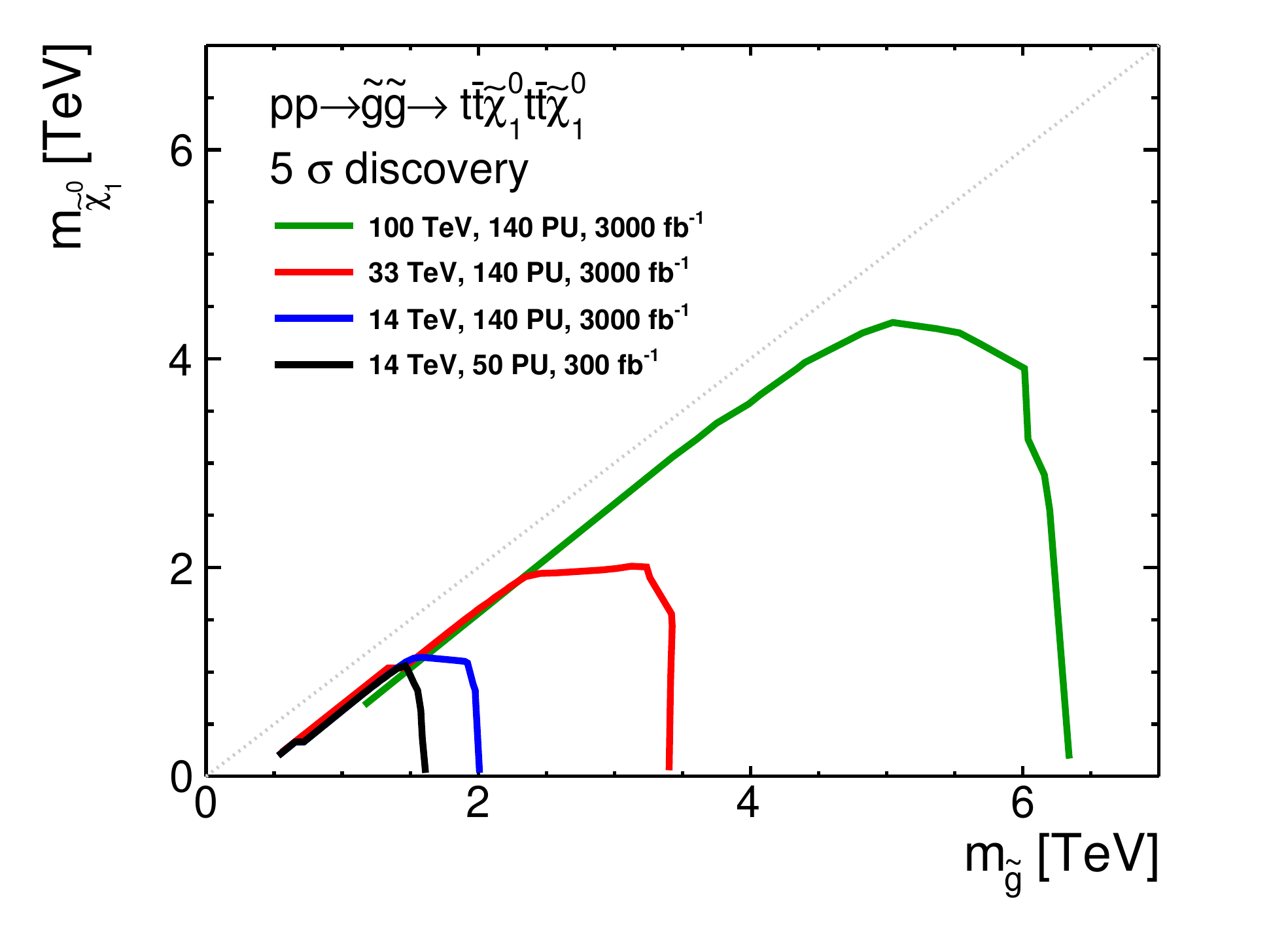}
\includegraphics[width=.48\columnwidth]{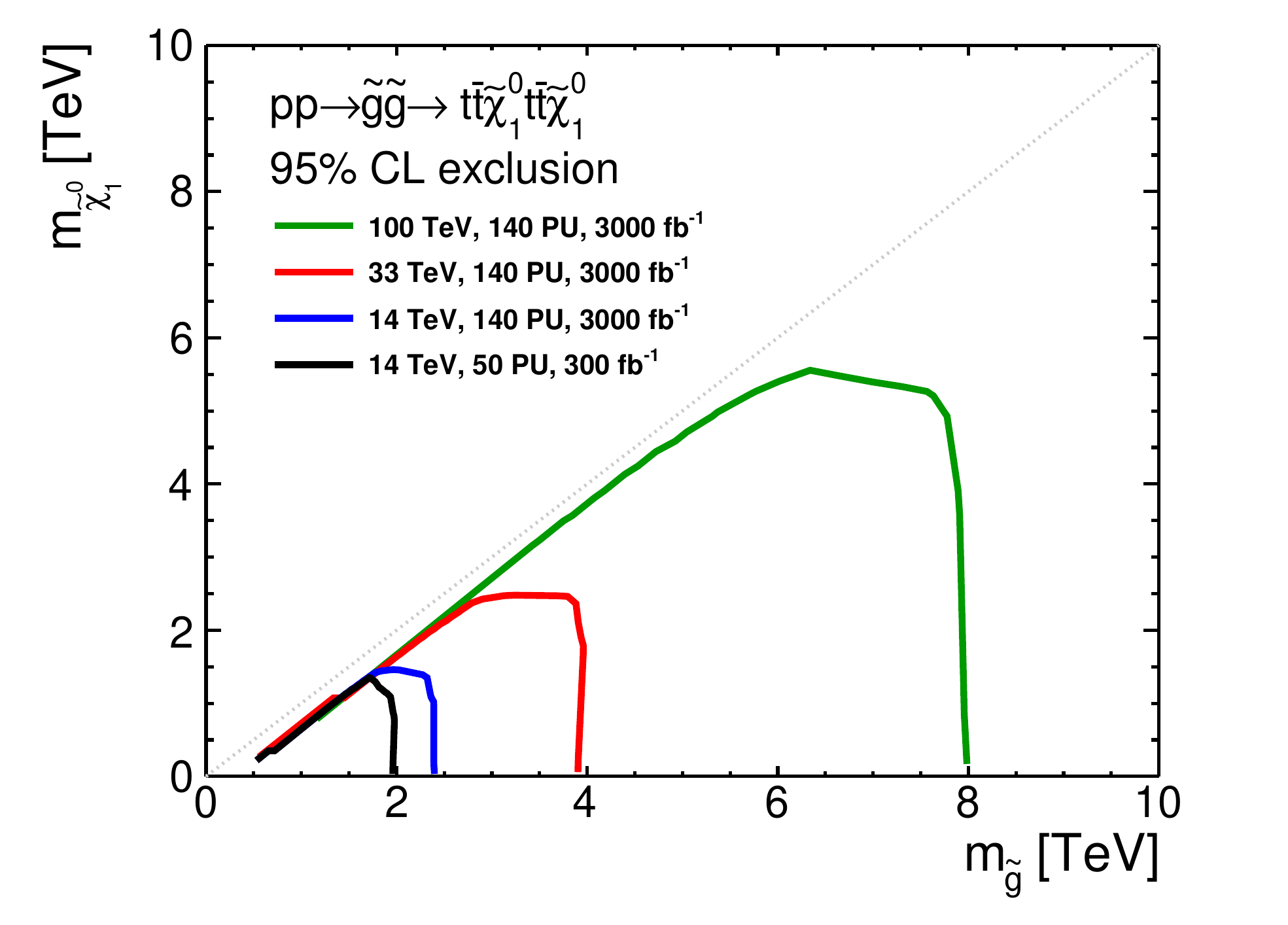}
  \caption{Results for the gluino-neutralino model with heavy flavor decays using the same-sign dilepton analysis strategy.  The left [right] panel shows the $5\,\sigma$ discovery reach [$95\%$ CL exclusion] for the four Snowmass collider scenarios.  A 20\% systematic error is assumed and pile-up is included.}
  \label{fig:GluinoNeutralino-HeavyFlavor}
\end{figure}

Results for the gluino-neutralino model with heavy flavor decays are shown in Fig.~\ref{fig:GluinoNeutralino-HeavyFlavor}.  The left [right] plot gives the $5\,\sigma$ discovery reach [$95\%$ CL exclusion].  For a massless neutralino, the $14 \TeV$ HL-LHC could discover a $\sim 2 \TeV$ gluino, a $33 \TeV$ proton collider could discover a $\sim 3.4 \TeV$ gluino, and a $100 \TeV$ proton collider could discover a $\sim 6.3 \TeV$ gluino.  Note that this analysis is also sensitive to compressed spectra due to dedicated signal regions; we estimate that models with $m_{\widetilde{g}} \simeq m_{\widetilde{\chi}_1^0} \sim 5.7 \TeV$ could be discovered with a 100 TeV machine.

\pagebreak
\section{Lessons Learned}
\label{sec:Lessons}
In this note, we have provided the discovery reach and expected limits for five simplified models at the four Snowmass proton collider scenarios.  Our focus was on models whose cross sections result from colored production.   These results provide a quantitative picture of what can be gleaned from potential luminosity and energy upgrades for the LHC.  For example, a $100$ TeV proton collider could discover an $11$ TeV gluino.  The remainder of this note is devoted to discussions of how the results depend on the assumptions about systematic errors and pile-up conditions made in this study.

When determining the reach of a given search, an optimization is performed in order to maximize significance.  All searches presented here assume a 20\% systematic error for the background prediction.  We find that the optimization procedure returns signal regions where systematic error dominates over the statistical uncertainty.  In the event of a discovery (or at least a strong hint of BSM physics), it is plausible that tremendous effort would be devoted to understanding and accounting for many systematic effects.  Given the number of events which would be available on tape, attempts to model backgrounds with data driven approaches should have minimal issues with statistics.  It is interesting to understand how much more could be learned from the data if these systematics are brought under control. In the gluino-neutralino model, fixing the luminosity and decreasing the systematic error from 20\% to 5\% increases the gluino limit by $300 \GeV$ for a massless neutralino.  

To assess the impact of the HL-LHC, it is especially interesting to understand the effect of a factor of 10 increase in luminosity at 14 TeV.  Note that the $300$ fb$^{-1}$ searches are ``systematics" limited; naively increasing the integrated luminosity will have no impact on the reach.  However, the cuts are re-optimized for the higher luminosity, and we find that the additional data allows harder cuts to be placed which yields an increase in the limit.  For gluino pair production with decoupled squarks, the cross section falls off by a factor of $\sim 10$ when $m_{\widetilde{g}}$ is varied from $2.0$ to $2.5 \TeV$.  A modest 500 GeV (25\%) improvement in expected limits is therefore the best case scenario for a background-free signal region. In practice, we find that the optimal signal regions are not background free and the reach improves more slowly --- a gluino could be discovered (assuming a massless neutralino) at 1.9 TeV with 300 fb$^{-1}$ and 2.3 TeV with a 3000 fb$^{-1}$ of integrated luminosity.  

Some of the results presented here are approaching the maximal possible masses that can be probed.  This can be estimated from the simple requirement that a non-zero number of signal events would be generated at a given energy and integrated luminosity.  For example, the gluino limit of 13.5 TeV for the 100 TeV collider corresponds to only $\sim 60$ SUSY events in a 3000 fb$^{-1}$ data set. This implies that this search is performing quite well, even in the presence of the non-trivial systematics.

In order to achieve integrated luminosities of $O(1000\text{ fb}^{-1})$, the machine configuration will be such that a large number of interactions per bunch crossing should be expected.  In order to understand the impact of pile-up on the results presented here, we reproduced the Monte Carlo samples including pile-up at the event-level and pile-up subtraction strategies at the reconstruction level  \cite{Anderson:2013kxz} for the $14$ TeV searches for the jets + \MET~and compressed spectrum strategies, and for all three energies for  the same-sign dilepton and single lepton searches.   We find that the results using our hadronic searches in Secs.~\ref{sec:GNLightFlavor}, \ref{sec:QN}, and \ref{sec:GluinoSquarkNeutralino} are effectively unchanged.  This statement holds for the $14$ TeV and $33$ TeV searches involving leptons as well.  However, with increase in energy the hardness of additional interactions increases, thus effecting lepton isolation. For the $100$ TeV collider lepton efficiencies are reduced since the average $p_T$ for a pile-up event has grown to the point that this contamination can deplete the number of isolated lepton leptons.  The reach of the $100$ TeV lepton based searches in Sec.~\ref{sec:GNHeavyFlavor} is reduced by $O(10\%)$ from these effects.  It is possible that some of the mass reach could be recovered if the lepton isolation criterion were adjusted for the $100$ TeV environment; exploring these issues is beyond the scope of this project.

In light of these additional lessons, our results provide robust projections for discovery reach and expected limits for the next phase of the LHC, and potential high energy proton colliders that could be built in the future.  In particular, the results here stand as some of the first quantitative estimates for the physics capabilities of a $100$ TeV proton collider using full background simulations.  These estimations clearly provide a strong physics case for the $14$ TeV LHC and beyond.

\pagebreak
\ack
TC is supported by the US DoE under contract number DE-AC02-76SF00515 and was supported in part by the NSF under Grant No. PHYS-1066293 while enjoying the hospitality of the Aspen Center for Physics.  AH is supported by the Alexander von Humboldt Foundation, Germany.  KH is supported by an NSF Graduate Research Fellowship under Grant number DGE-0645962.  SP was supported in part by the Department of Energy grant DE-FG02-90ER40546 and the FNAL LPC Fellowship.  JGW is supported by the US DoE under contract number DE-AC02-76SF00515. The authors are supported by grants from the Department of Energy Office of Science, the Alfred P. Sloan Foundation and the Research Corporation for Science Advancement.

\footnotesize
\setstretch{1.0}
\bibliography{SimplifiedModels}
\end{document}